# Realization of a vertical topological p-n junction in epitaxial $Sb_2Te_3$ / $Bi_2Te_3$ heterostructures

Markus Eschbach[1], Ewa Młyńczak[1,6] , Jens Kellner[7] , Jörn Kampmeier[2] , Martin Lanius[2] , Elmar Neumann[8] , Christian Weyrich[2] , Mathias Gehlmann[1] , Pika Gospodarič[1] , Sven Döring[1] , Gregor Mussler[2] , Nataliya Demarina[3] , Martina Luysberg[5,8] , Gustav Bihlmayer[4] , Thomas Schäpers[2] , Lukasz Plucinski[1] , Stefan Blügel[4] , Markus Morgenstern[7] , Claus M. Schneider[1] , and Detlev Grützmacher[2]

1 Forschungszentrum Jülich GmbH, Peter Grünberg Institute (PGI-6) and JARA-FIT, 52425 Jülich, Germany
2 Forschungszentrum Jülich GmbH, Peter Grünberg Institute (PGI-9) and JARA-FIT, 52425 Jülich, Germany
3 Forschungszentrum Jülich GmbH, Peter Grünberg Institute (PGI-2) and JARA-FIT, 52425 Jülich, Germany
4 Forschungszentrum Jülich GmbH, Peter Grünberg Institute (PGI-1) and JARA-FIT, 52425 Jülich, Germany
5 Forschungszentrum Jülich GmbH, Peter Grünberg Institute (PGI-5) and JARA-FIT, 52425 Jülich, Germany
6 Faculty of Physics and Applied Computer Science, AGH University of Science and Technology, al. Mickiewicza 30, 30-059 Kraków, Poland
7 RWTH Aachen University, II. Institute of Physics B and JARA-FIT, 52074 Aachen, Germany
8 Forschungszentrum Jülich GmbH, Ernst Ruska-Centrum for Microscopy and Spectroscopy with Electrons, 52425 Jülich, Germany

Correspondence and requests should be addressed to L.P. (email: l.plucinski@fz-juelich.de)

# Abstract

3D topological insulators are a new state of quantum matter which exhibits both a bulk band structure with an insulating energy gap as well as metallic spin-polarized Dirac fermion states when interfaced with a topologically trivial material. There have been various attempts to tune the Dirac point to a desired energetic position for exploring its unusual quantum properties. Here we show a direct experimental proof by angle-resolved photoemission of the realization of a vertical topological p-n junction made of a heterostructure of two different binary 3D TI materials $Bi_2Te_3$ and $Sb_2Te_3$ epitaxially grown on Si(111). We demonstrate that the chemical potential is tunable by about 200 meV when decreasing the upper $Sb_2Te_3$ layer thickness from 25 to 6 quintuple layers without applying any external bias. These results make it realistic to observe the topological exciton condensate and pave the way of exploring other exotic quantum phenomena in the near future.

It is well known that binary compounds $X_2Y_3$ based on high-Z elements with sufficiently strong spin-orbit coupling, such as Bi, Te, Se and Sb, belong to the class of strong 3D topological insulators (TIs). They host an odd number of gapless Dirac cone-like linear dispersive surface states with chiral spin-momentum locking, which are located around high symmetry points in the surface Brillouin zone[1-5]. From a practical point of view, these materials can be a basis for novel dissipationless spintronic devices because the propagation direction of their surface electrons is robustly locked to their spin-orientation, i.e. back-scattering in charge transport is prohibited as long as time-reversal symmetry is preserved[6,7].

Studying the Dirac fermion states promises the verification of exotic quantum phenomena, such as the image magnetic monopole which may exist in TIs due to proximity effects to a ferromagnetic material[8] or Majorana fermions which can be induced at the interface of a strong TI and an s-wave superconductor[9]. More specifically, the spatial separation of variable Dirac cone structures opens up the possibility to study what was proposed as the horizontal topological p-n junction[10], where the effect of the lateral variation of the chemical potential on the spin-locked transport can be investigated, including its control by external electric fields[11].

Moreover, the vertical (with respect to the sample plane) separation of the Dirac cones might enable the observation of the so-called topological exciton condensate which is proposed to exhibit fractionally charged excitations (similar to Majorana fermions) in its vortices without any additional interface[12]. The only prerequisites are separated electron- and hole-type Dirac fermions on opposite surfaces which interact electrostatically.

There has been considerable amount of research to precisely tune the position of the Fermi level $E_F$ in topological Dirac cones. Firstly, this can be achieved by surface doping[13] which, however, does not lead to a suppression of the bulk conductivity. Another successful path is to gradually tune the composition in a ternary[11,14-16] (or even quaternary[16]) alloy, like $(Bi_{1-x}Sb_x)_2Te_3$. Since typical epitaxially-grown layers of $Bi_2Te_3$ ($Sb_2Te_3$) turn out to be of n-(p-) type charge character which are dominated by electron (hole) transport in the bulk[17], this alloying leads to an effective compensation of charge and thus to a shift of the chemical potential and tunable surface states and eventually also suppression of the bulk conductivity.

Similarly, bringing together two different binary TI films to create a vertical topological p-n junction should also lead to compensation of charge within the depletion layer formed at their interface. However, the effect of such a topological p-n junction on the topologically protected surface states or the surface electronic structure in general has not been reported so far.

In this article we report on the direct observation of thickness dependent electronic shifts of the chemical potential in vertical topological p-n junctions by means of angle-resolved photoemission spectroscopy (ARPES). The junctions are created in $Sb_2Te_3$ / $Bi_2Te_3$ heterostructures of variable layer thickness, grown by molecular beam epitaxy which assures high crystalline quality and high accuracy of the thickness and composition of the thin films. The thickness of the underlying $Bi_2Te_3$ layer is kept constant for all samples as ~ 6 quintuple layers (QLs), whereas the $Sb_2Te_3$ layer thickness $t$ is varied $t$ = 25, 15, 7, 6 QL. Modifying the top layer thickness results in a varying influence from the buried layer on the probed upper surface and thus, to a Dirac point shifting up to 200 meV with respect to the Fermi level. In this way we are able to alter a $Sb_2Te_3$ surface from being of p-type charge carrier character to n-type by reducing the thickness above the $Bi_2Te_3$ layer.

Thus, we believe that our findings add a fundamentally new approach to the conventional ones, such as doping or biasing, to engineer the band structure in TIs and especially their Dirac cone by intrinsic interfacial effects. Further, we believe that such advanced synthesis techniques will allow for the study of the interaction of opposing Dirac cones and potential new quantum states such as the topological exciton condensate.

# Results

**Structural analysis.** High-resolution scanning transmission electron microscopy (STEM) measurements were carried out in order to investigate structure and quality of the heterostructures. Figure 1a displays a high-angular annular dark field (HAADF) image of a 15 QL $Sb_2Te_3$ / 6 QL $Bi_2Te_3$ sample. According to the difference in atomic number, Bi atomic columns appear brightest. The crystalline quality and the degree of structural order of the individual quintuple layers, which are clearly separated by van der Waals gaps, are very high. Only at the interface to the Si substrate, the contrast is slightly deteriorated due to amorphisation during preparation of the sample for STEM (see methods section) or the additional Te bilayer at the Si interface which was reported by Borisova et al.[18]. This contrast change is also observed in very thin areas of the sample. Hence, only thick areas are suitable for investigation which implies a loss of resolution. Nonetheless, the individual atomic columns are clearly revealed, which is highlighted in the inset displaying four quintuple layers across the interface at higher magnification with a structural model as overlay. In Fig. 1b the intensity averaged within the red frame (in Fig. 1a) is plotted versus distance (also serving as scale of the STEM image in a). Towards the Si substrate a decrease in counts is observed indicating a reduction in specimen thickness, which is in line with the observed amorphisation. At the $Sb_2Te_3$ / $Bi_2Te_3$ interface the intensity is observed to decrease over a region of two quintuple layers, i.e. 2 nm. Since the contrast within the $Sb_2Te_3$ remains constant, we assume a constant thickness across the interface as well. Hence, the intensity gradient across the small interface region implies intermixing of Bi and Sb. Since STEM is a very local probe, additional characterization, such as low energy electron diffraction (LEED) and Auger electron spectroscopy (AES) depth profiling, were performed. Supplementary Note 1 gives detailed information on AES depth profiling of the 15 QL $Sb_2Te_3$ / 6 QL $Bi_2Te_3$ sample, being in good agreement with our STEM data. LEED reveals the well-known sixfold diffraction pattern indicating high crystalline perfection of the top surface (Supplementary Fig. 1c).

**Transport measurements.** High quality MBE-grown $Bi_2Te_3$ films are known to exhibit mostly n-type charge carriers due to Te vacancies that introduce donors. Besides vacancies, also ionized $Bi_{Te}$ antisite defects generate the n-type doping[17,19]. On the contrary, in $Sb_2Te_3$ the major defects are $Sb_{Te}$ antisite defects which impose p-type charge carriers. For this reason, $Sb_2Te_3$ / $Bi_2Te_3$ heterostructures are expected to exhibit a separation of opposite carrier character making them a natural p-n junction system. To prove the existence of different regimes of charge carrier types we performed magnetic field-dependent transport measurements at 1.4 K in standard Hall-bar geometry with a sample width between 20 and 40 µm. The resulting Hall resistances are shown in Fig. 2.

We observe a transition from n- to p-type regime for increasing top $Sb_2Te_3$ layer thickness. The slope of the transversal Hall resistivity $R_{xy}$ changes from negative to positive between a sample thickness of 6 QL (green curve) and 17 QL $Sb_2Te_3$ (red curve). This implies that the electronic transport is mostly dominated by hole (p-type) and electron (n-type) transport, respectively. There exists a non-linearity of the Hall resistance at lower fields which is interpreted as an effect due to the coexistence of both p- and n-type charge carriers in the samples. However, it is neither as strong nor it changes its slope as it was reported for certain gate voltages previously in similar heterostructures[20]. Furthermore, the slight change of the slope between the green (6 QL sample) and black curve (3 QL sample), which is opposite to the general trend, is due to the accuracy of this measurement.

**Electronic structure.** The detailed surface electronic structure of the studied heterostructures was mapped using high-resolution ARPES. Figure 3 displays long scale $E_B$ vs. $k_\parallel$ ARPES maps along trajectories traversing the $\bar{\Gamma}$-point of the surface Brillouin zone recorded with $h\nu = 21.22$ eV. The exact cut directions in $k$-space were deduced by Fermi surface mapping and are highlighted in the insets of Fig. 4a,b,c,d, and e. The plotted overview spectra all show dispersing bulk bands at relatively low background intensity, which signals the high crystalline quality of the samples. Typical features of the $Sb_2Te_3$ band structure[21] are revealed, such as the prominent Rashba-split surface state located between $E_B = 0.4 - 0.8$ eV and $k_\parallel = \pm 0.28$ Å$^{-1}$ in a spin-orbit induced gap within the projected band structure[22]. This feature is identified for all heterostructures. Furthermore, indications of the topologically protected Dirac cone states near the Fermi level are found in each spectrum. The photoemission cross section for these states is small at 21.22 eV, however, they can be analyzed in detail with lower photon energy (see next paragraph). An *ab initio* calculated electronic structure of a 6 QL thick $Sb_2Te_3$ film along the corresponding crystallographic direction was superimposed on each spectrum to confirm the origin of the spectral features[21].

The fact that all maps in Fig. 3 exhibit similar spectral features which originate from pure $Sb_2Te_3$ allows to determine the energetic shift of the entire band structure towards higher binding energies for decreasing $Sb_2Te_3$ layer thickness. As highlighted by the respective energy distribution curves in Fig. 3, the bottom of the prominent Rashba-split surface state at $\bar{\Gamma}$, which has the largest spectral weight in these spectra, shifts by about 250 meV from 25 QL to 6 QL $Sb_2Te_3$.

The same effect of a shifting band structure is observed for the Dirac cone-like topological surface band near the Fermi level. Figures 4a,b,c,d, and e present the Fermi surface $k_x$ vs. $k_y$ maps and Fig. 4f,g,h,i, and j the $E_B$ vs. $k_\parallel$ spectra from a region close to the Fermi level for the pure $Sb_2Te_3$ film and the set of heterostructures (25 QL, 15 QL, 7 QL, and 6 QL, respectively) measured at $h\nu = 8.44$ eV. The magnified calculated electronic structure is superimposed in each spectrum. The insets in Fig. 4a,b,c,d, and e illustrate the exact cut direction through the surface Brillouin zone. Additionally, Fig. 4k,l,m,n, and o depict the corresponding momentum distribution curves.

In each of these high-resolution ARPES spectra the Dirac cone can be observed. The spectra reveal that for decreasing $Sb_2Te_3$ top layer thickness the chemical potential of the sample surface is shifted from within the valence band through the forbidden band gap and towards the conduction band. Thereby, the Dirac point (DP) crosses the Fermi level at about 15 QL $Sb_2Te_3$.

The samples consisting of pure $Sb_2Te_3$ (a,f,k) and the heterostructures with 25 QL (b,g,l) and 15 QL (c,h,m) $Sb_2Te_3$ top layer exhibit a Fermi level which still cuts the valence band. This is visible as sizable spectral weight from bulk bands with hexagonal symmetry within the Fermi surfaces. On the contrary, for the two thin films with 7 QL (d,i,n) and 6 QL (e,j,o) top layer thickness the Fermi level is well above the valence band and apparently inside the fundamental band gap with a Dirac point below $E_F$. This is in perfect agreement with the transport data shown in Fig. 2.

From the superimposed calculated electronic structure the position of the Dirac point with respect to the Fermi level is determined with an accuracy of $\pm 20$ meV. This method is known to be more precise than determining the intersection of two regression lines fitted to the Dirac cone[21]. The extracted binding energy positions of the Dirac points $E_B(DP)$ are listed in Table 1. The total energetic shift deduced from the shifting Dirac point from the thickest to the thinnest heterostructure sample is about 200 meV, albeit slightly lower than the number derived from the wide energy spectra in Fig. 3. Furthermore, the Fermi velocity $v_F$, derived from linear fits to the topological surface states close to the Fermi level according to $E_B = \hbar v_F |k_\parallel|$, is given in Table 1. Similar values have been obtained for pure $Sb_2Te_3$ and $Bi_2Te_3$ thin films in previous works[11,10].

The observed energetic shifts of the electronic structure in ARPES and transport are highly reproducible and in accordance with the expected charge carrier character of the sample surface which is of p-type or n-type depending on the distance to the $Bi_2Te_3$ layer.

**Comparison with 1D Schrödinger-Poisson model.** Although there exist theoretical models which treat 3D TIs as highly doped narrow band gap semiconductors and question their ultimate bulk resistivity due to poorly screened random potential fluctuations[23,24], recent experiments indicate that these limitations can be overcome in certain TI compounds[25] or by the use of dual-gating of thin films[26,27]. Furthermore, Brahlek et al.[28] show that band bending effects indeed can lead to bulk insulating states in the Mott sense.

The latter makes us confident to be able to compare our results with a simulated band profile by modeling the TI heterostructure system for various thicknesses of the $Sb_2Te_3$ layer and self consistently solving 1D Schrödinger and 1D Poisson equations. The results of these simulations are shown in Fig. 5. Good agreement to the experimental data, considering the energetic position of the valence band maximum with respect to the Fermi level (Fig. 5a), is found if one assumes a donor-type (acceptor-type) charge carrier density of $2 \cdot 10^{19}$ cm$^{-3}$ ($-2 \cdot 10^{18}$ cm$^{-3}$) in the Bi$_2$Te$_3$ (Sb$_2$Te$_3$) layer and an additional charge of $1 \cdot 10^{12}$ cm$^{-2}$ at the Bi$_2$Te$_3$ / Si interface and the Sb$_2$Te$_3$ surface (all details of the model are given in Supplementary Note 2). To account for our knowledge of the slight intermixing at the interface we also include into the model a 5 nm wide interface region where the charge carrier density follows an experimentally deduced profile (by AES depth profiling, see Supplementary Fig. 1), which lead to further improvement of the comparison. Figures 5b and c exemplarily show the band diagram of valence and conduction band throughout the entire system for Sb$_2$Te$_3$ thicknesses of $x = 35$ nm and $x = 10$ nm, respectively. In the thin top layer regime, one can see that the samples are in an insulating state with the conduction band minimum close to the Fermi level, whereas for the thick top layers the Fermi level cuts well through the valence band.

Finally, Fig. 5a summarizes the resulting position of the valence band maximum at the surface to vacuum plotted versus Sb$_2$Te$_3$ layer thickness. The dashed black line marks the Fermi level. The conduction band edge is not shown but would follow the same slope. Additionally, the figure depicts the experimentally deduced values from the ARPES measurements (red dots). Experimentally, we observe a shift of the entire electronic structure of about 200 meV, while the simulations would predict a larger shift of about 350 meV. However, the trend is reproduced and thus, the feasible agreement between our 1D model and our ARPES data again confirms that we have created a topological p-n junction.

# Conclusion

In conclusion, we have presented a new reliable way to precisely and robustly tune the charge carrier character and the chemical potential in 3D topological insulators by the creation of built-in interfacial electric fields in a vertical topological p-n junction. This method does not introduce disorder due to doping and does not require any external bias. Our heterostructures are a convenient playground for the study of Majorana fermions, which are predicted to be observed in tunneling spectroscopy measurements of 3D TI-superconductor interfaces, if the Dirac point is adjusted precisely at the Fermi level[9,29]. Moreover, Dirac electrons which reside in the (n-type) topological state at the interface between Bi$_2$Te$_3$ and silicon (which can not be probed by ARPES) can couple to the (p-type) surface Dirac electrons such that a topological exciton condensate with similar properties as a topological superconductor is formed[12]. This new state of correlated quantum matter is so far elusive, but opens up a feasible alternative to probe effects like charge fractionalization in vortices or topological magnetoelectric effects.

In this respect, compared to dual-gating of a single TI layer, our approach of combining two binary TI layers in a p-n junction is much more versatile and, moreover, has the advantage of linearly dispersing Dirac states on both sides of the heterostructure which enables the existence of two identical electron and hole Fermi surfaces. Consequently, the built-in spatial asymmetry of the Dirac bands and the reliable tunability of the chemical potential by manipulating purely internal structural parameters pave the way for studying exciting novel phenomena with potential applications in spintronics.

# Methods

**Growth.** A set of epitaxial Bi$_2$Te$_3$/Sb$_2$Te$_3$ bilayers was grown by means of molecular beam epitaxy on high ohmic n-type doped Si:P(111) substrates of $10 \times 10$ mm size (doping level $\sim 10^{13}$ cm$^{-3}$) under ultra-high vacuum conditions. Bi$_2$Te$_3$ is known to grow epitaxially on Si:P(111), forming films of high structural quality[30]. It grows in a rhombohedral structure with five atomic layers, known as quintuple layers (QL), as a basic unit forming relatively strong covalent bonds between the atoms within one QL, whereas the interaction among the consecutive QLs is of the van der Waals type. Since $Bi_2Te_3$ and $Sb_2Te_3$ have very similar lattice constants ($Bi_2Te_3$: $a = 4.385$ Å, $c = 30.49$ Å; $Sb_2Te_3$: $a = 4.264$ Å, $c = 30.458$ Å), $Sb_2Te_3$ grows epitaxially on $Bi_2Te_3$ as well. The growth rates were kept constant at $v_{Bi_2Te_3} = 11$ nm/h and $v_{Sb_2Te_3} = 9$ nm/h.

In addition to the set of $Sb_2Te_3$ / $Bi_2Te_3$ bilayers, a 10 QL thick $Sb_2Te_3$ film was prepared to serve as a reference sample. In those thickness regimes the coupling and hybridization of adjacent surface states through the layers can be neglected[31,32]. The detailed growth parameters, structural analysis and characterization of the system have been presented earlier[30,18].

**ARPES.** After growth the samples were transferred under ambient conditions into the high-resolution ARPES apparatus (base pressure $< 1 \cdot 10^{-10}$ mbar). Subsequently, they were annealed up to 220-250°C for two minutes to desorb surface contaminations. Afterwards the samples were cooled down and all measurements were carried out at approximately 15 K. The spectra were taken with an MBS A1 analyzer set to an energy resolution of 10 meV for all presented ARPES data. The angular resolution in these ARPES experiments is $< 0.4°$. Mapping of the electronic structure is achieved by rotating the sample around the axis which is aligned along the long axis of the analyzer entrance slit.

Two different photon energies were used in these studies. To obtain overview spectra, non-monochromatized He Iα resonance radiation of $h\nu = 21.22$ eV was employed since it allows access to the entire valence band (Fig. 3). Taking advantage of the large photoemission cross section of $Sb_2Te_3$ surface states for low energy photons[21], monochromatized light from a microwave-driven Xenon discharge lamp with $h\nu = 8.44$ eV (Fig. 4) was used for the detailed analysis of the surface states.

**High-resolution STEM.** Structural investigations on the atomic scale have been performed with an aberration corrected scanning transmission electron microscope (FEI Titan 80-300) on cross-sectional specimen. The contrast of the high-angular annular dark field (HAADF) images approximately scales with the atomic number $Z^2$ which allows to distinguish between elements of large difference in atomic number, such as Sb and Bi. Specimens have been prepared by focused ion beam etching using first 30 keV Ga ions followed by a 5 keV final treatment. Ar ion milling with the NanoMill operated at 900 V and subsequently at 500 V was employed to reduce the surface damage introduced by FIB.

**DFT calculations.** All ARPES spectra presented here are compared with theoretical calculations of the electronic band structure of a 6 QL thick $Sb_2Te_3$ film. The theoretical spectra were calculated by full-relativistic density functional theory using the generalized gradient approximation and a full-potential linearized augmented plane wave (FLAPW) method which is implemented in the FLEUR code. For more details see ref. 33.

**1D Schrödinger-Poisson model.** See Supplementary Note 2.

# Acknowledgments

We highly acknowledge fruitful discussions with Prof. Lüth and the extraordinary technical support along with the design of the ARPES apparatus by Bernd Küpper. This work was performed within the SPP 1666 programme in collaboration with the Virtual Institute for Topological Insulators (VITI), which is financially supported by the Helmholtz Association.

# Author contributions

M.E., E.M., Je.Ke., M.G., S.D., and P.G. carried out all ARPES and AES depth profiling experiments under the supervision of L.P. and C.M.S.

Jö.Ka. and G.M. grew the sample via MBE. C.W. and Th.S. performed electrical transport measurements after the ARPES measurements.

M.La., E.N., and M.Lu. prepared the FIB lamellas and performed the HR STEM measurements.

N.D. simulated the sample system in a 1D Schrödinger-Poisson model. G.B. and S.B. provided the ab initio DFT band structure calculations for exact comparison to the ARPES data.

M.E., E.M., and L.P. wrote the paper with contributions from all co-authors. D.G. initiated the project which was supervised by M.M., D.G., and C.M.S.

# Figures

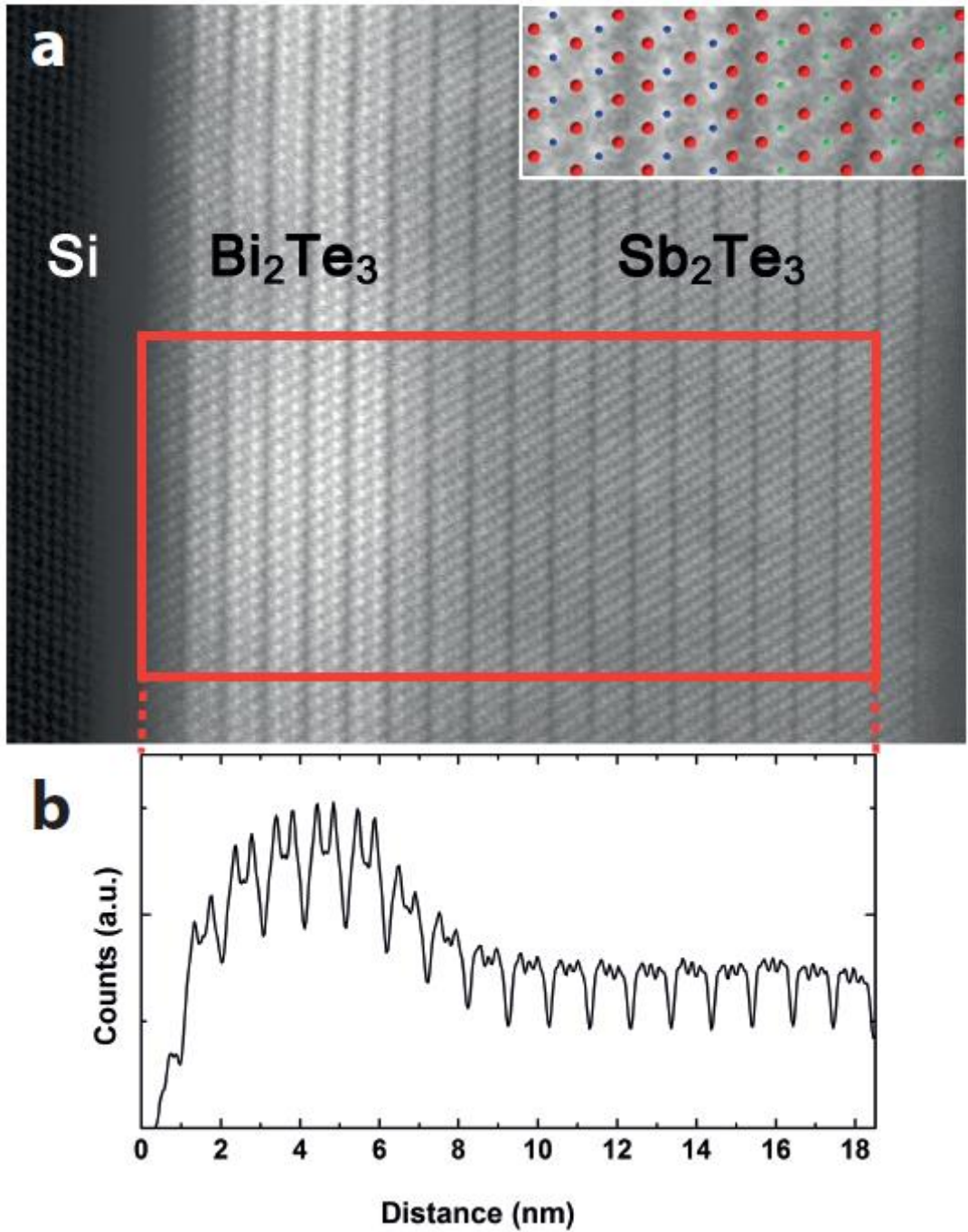

**Figure 1 | Structural analysis of the 15 QL $Sb_2Te_3$ / 6 QL $Bi_2Te_3$ sample via STEM. (a)** HAADF image of atomic resolution. The large overview image reveals the high quality of the crystal. Van der Waals separated quintuple layers can be observed. The contrast in the image is related to the size of the atoms on which electrons are scattered, i.e. chemical contrast is obtained. To estimate the size of the intermixed interface region a line profile is plotted in **(b)** integrated over the red rectangle in **a**. This line profile also serves as a scale bar for **a**. The inset in **a** shows a magnified region across the interface of the two layers with a structural model superimposed (blue atoms = Bi; green atoms = Sb and red atoms = Te).

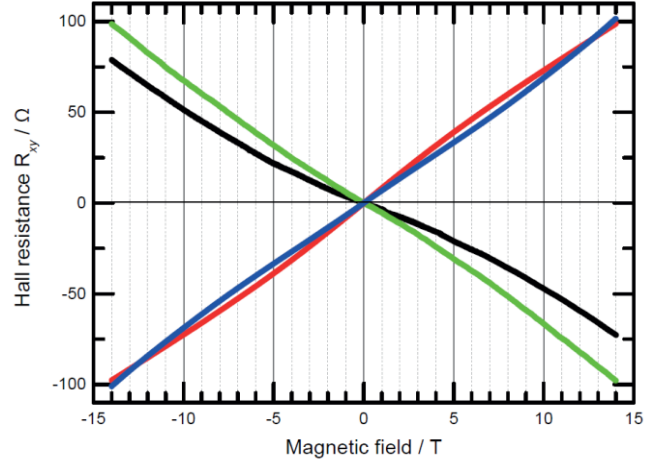

**Figure 2 | Field-dependent transport measurements.** Hall resistance $R_{xy}$ of 4 different samples with varying top $Sb_2Te_3$ layer thickness investigated at fixed gate voltages and low temperature. For thinner top layer thickness of 3 QL (black curve) and 6 QL (green curve) the heterostructure is in an n-type (electron) transport regime, whereas for thicker films of 17 QL (red curve) and 28 QL (blue curve) p-type (hole) transport is dominant.

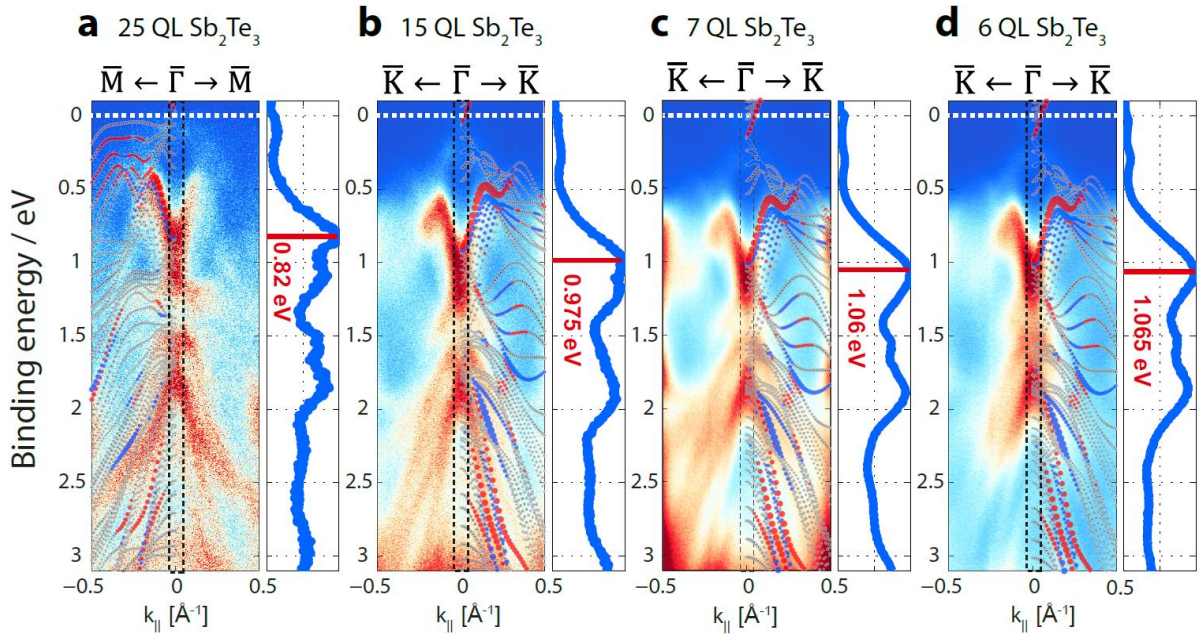

**Figure 3 | Wide energy $E_B$ vs. $k_\parallel$ ARPES maps.** 25 QL (a), 15 QL (b), 7 QL (c) and 6 QL (d) $Sb_2Te_3$ samples measured along indicated crystallographic directions using $h\nu = 21.22$ eV. The electronic structure of a 6 QL thick $Sb_2Te_3$ slab calculated by DFT along the corresponding crystallographic direction is superimposed. Red and blue dots in this calculation refer to opposite in-plane spin orientation. The Fermi level is indicated by the white dashed line. The energy distribution curves (EDCs) which are integrated over the black dashed area are shown on the right of each ARPES map and mark the energetic position of the most prominent features. The main feature being the bottom of the lower Rashba-split surface state serves as a gauge for the observed energetic shift.

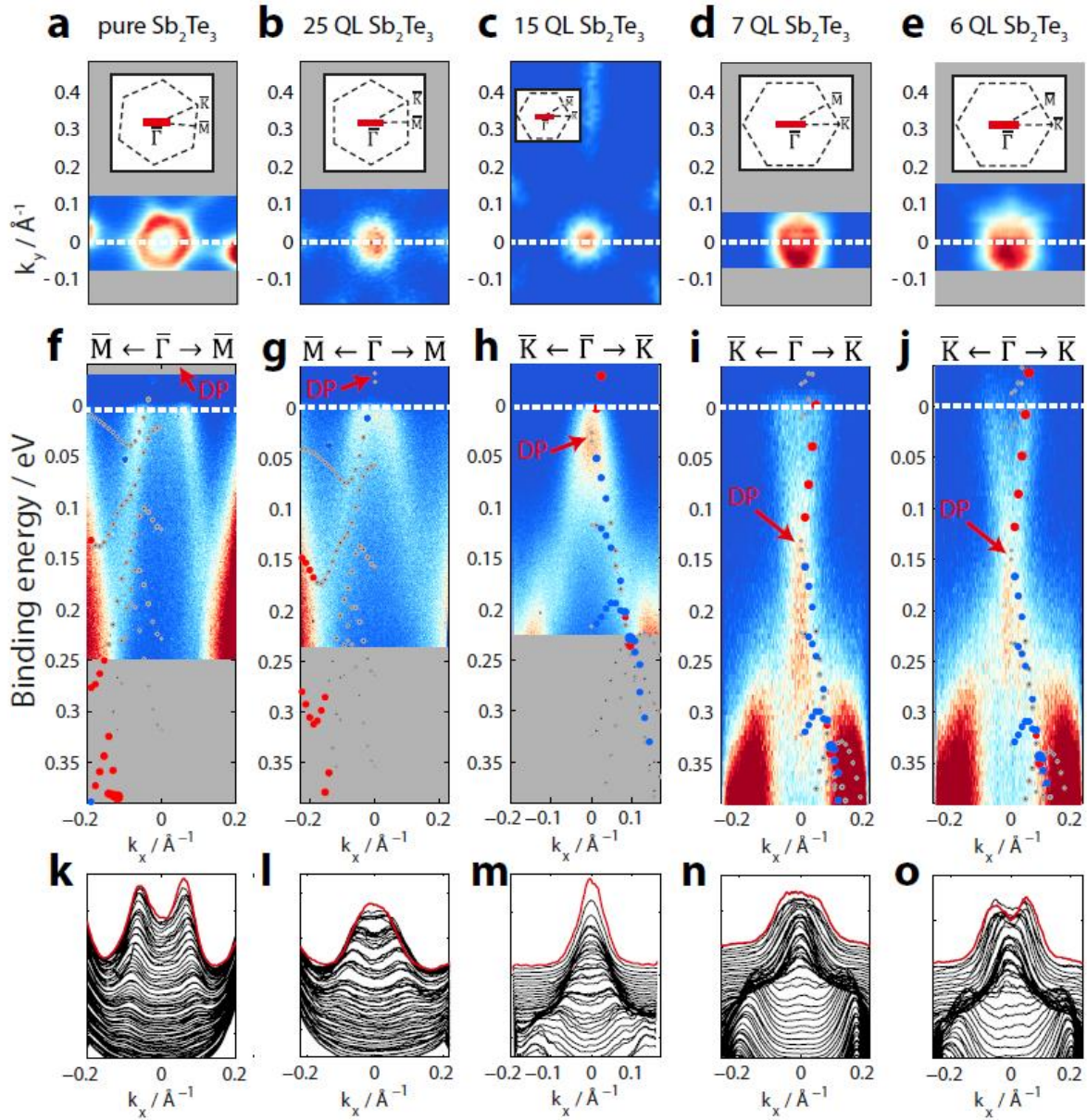

**Figure 4 | High-resolution ARPES close to the Fermi level using $h\nu = 8.44$ eV. (a,f,k)** present the results obtained for the reference single $Sb_2Te_3$ film. For the heterostructures the $Sb_2Te_3$ top layer thickness is marked on top. **(a,b,c,d,e)** depicts the measured Fermi surface maps $k_x$ vs. $k_y$ for $E_B = E_F$. The black dashed lines in the insets depict the hexagonal shape of the surface Brillouin zone. This symmetry character is also conserved for the shape of the surface state as one departs from the Dirac point. The white dashed lines (= red line in the inset) indicate the cut direction where the corresponding normal emission spectra **(f,g,h,i,j)** were recorded. The Dirac point is marked by red arrows and the band structure calculations from DFT with adopted Fermi energy are superimposed in each spectrum. Again, red and blue dots here represent opposite in-plane spin polarization of the states. Plots **(k,l,m,n,o)** show the respective momentum distribution curves at binding energies from $E_B = 0.2$ eV (bottom) to $E_B = E_F = 0$ eV (top, marked by the red line) of the spectra above.

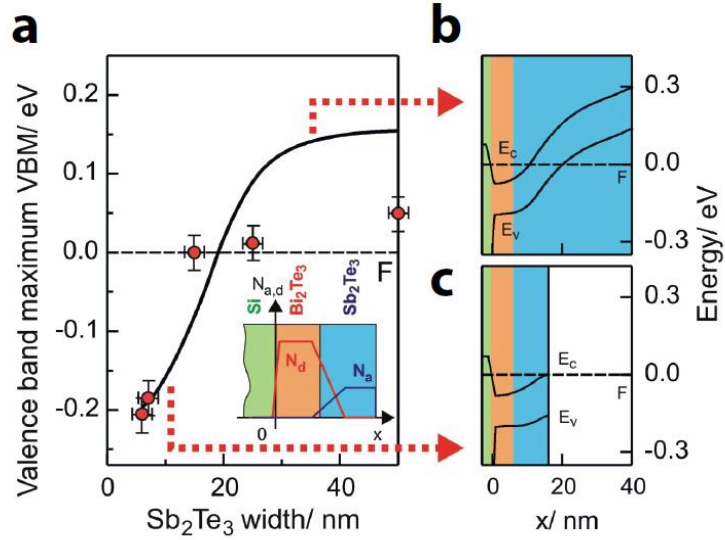

**Figure 5 | Result of the 1D model using Schrödinger-Poisson equation. (a)** Calculated energetic position of the valence band maximum with respect to the Fermi level at the surface to vacuum for different $Sb_2Te_3$ layer thicknesses (black line) and experimentally derived values of the VBM from ARPES (red dots). A large error or ± 25 meV was estimated on the position of the VBM because determination from ARPES is difficult. An error of ± 2 nm was assumed on the accuracy of the film thickness from combined XRR and TEM investigations. The model assumes the creation of a depletion layer which causes band bending. **(b)** and **(c)** show the band diagram of both valence and conduction band throughout the entire system for top layer thicknesses of 35 nm and 10 nm, respectively (connected by red dashed arrows to the curve in **a**). Green is the Si substrate, orange is $Bi_2Te_3$ and blue $Sb_2Te_3$.

| Sample | $E_B(DP)$ / meV | $v_F$ / $10^5$ ms$^{-1}$ |
|---|---|---|
| Pure $Sb_2Te_3$ | - 65 | 4.4 |
| 25 QL | - 35 | 2.5 |
| 15 QL | + 30 | 2.2 |
| 7 QL | + 140 | 5.2 |
| 6 QL | + 145 | 4.8 |

**Table 1 | Dirac point binding energies and Fermi velocities of the different samples.** Binding energy position of the Dirac point is extracted from the superimposed calculations. Negative (positive) binding energy refers to the unoccupied (occupied) band structure above (below) the Fermi level. Third column shows the Fermi velocities derived from linear fits to the surface states close to the Fermi level.

# Supplementary Information

## Supplementary Note 1

**Composition profiling.** Selected samples have been investigated by repeated gentle 500 eV $Ar^+$ ion sputtering and subsequent Auger electron spectroscopy (AES) cycles in order to determine the composition profile. Supplementary Fig. 2 shows the result of such AES depth profiling for a 15 QL $Sb_2Te_3$ / 6 QL $Bi_2Te_3$ sample. From the single AES spectra presented in Supplementary Fig. 1a the relative amount of Bi, Te, Sb and Si was determined. Hence, the composition can be plotted against the sputtering cycle and, if one uses the known sample thickness from STEM (see Fig. 1), this can be recalculated into a real film thickness, as done in Supplementary Fig. 1b. Separated regions of $Sb_2Te_3$ and $Bi_2Te_3$ can be clearly distinguished but again a sizable diffusion of Sb and Bi at the interface is found which might have been induced by the additional annealing step. To be more precise, firstly, Sb diffuses to the $Bi_2Te_3$ − Si interface and, secondly, Bi shows a non-vanishing signal throughout the entire heterostructure. However, at the surface to vacuum the characteristic low energy 103 eV NOO Auger peak of Bi was found to be small in all samples. The depth profiling confirms the existence of an intermixed interface region between the two TI layers of width of a few nm. The real extension of this region will be smaller than the red or blue profiles because one has to take several broadening effects into account.

Firstly, the resulting AES signal will be a convolution of the true profile and a thickness-dependent function that includes inelastic mean free paths of the Auger electrons (simulated by the dashed lines in Supplementary Fig. 1b). Such a function gives an upper limit of the accuracy of the AES profiling method, assuming that the sputtering process proceeds homogeneously, i.e. in the layer-by-layer mode. Secondly, the measured profile is convoluted with the depth resolution function [1], taking into account sputtering-induced changes in the composition and surface roughness, the assessment of which is beyond the scope of the present analysis. This means that the true profile is expected to be steeper than derived here, which is in good agreement with the presented STEM data and also means that our cleaning procedure did not significantly influence the quality of the interface.

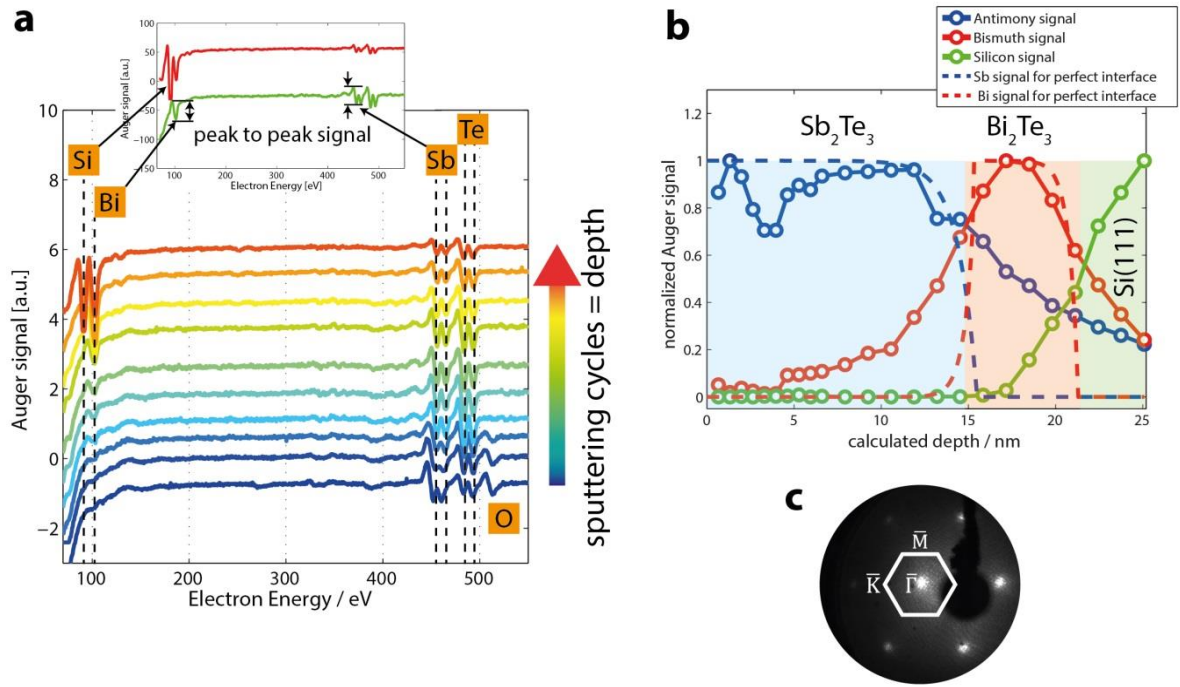

**Supplementary Figure 1 | Structural analysis of the 15 QL $Sb_2Te_3$ / 6 QL $Bi_2Te_3$ sample via AES depth profiling. (a)** Single AES spectra obtained after repeated cycles of ion sputtering (from blue = untreated to red). Peaks belonging to different elements are marked. The inset shows how the peak-to-peak signal was obtained. **(b)** The peak-to-peak signal from Bi (red), Sb (blue) and Si (green) is plotted against sputtering time which is recalculated into sample thickness by calibration via the STEM measurements. The dashed lines mark the lower limit of accuracy which this method can provide assuming an ideal sharp interface between the two materials. **(c)** Low-energy electron diffraction (LEED) pattern for this sample showing the high crystalline quality of the (111) surface.

## Supplementary Note 2

**1D Schrödinger-Poisson model.** To estimate the band bending within the p-n junction created by the connection of two narrow band gap semiconductors with opposite dominating charge character, we modeled the system in 1D solving the conventional 1D Poisson and 1D Schrödinger equations self consistently [2]. The Schrödinger equation was written for the envelope function using the effective mass approximation. Numerically, both equations were iteratively solved and the solution was altered until the charge neutrality of the structure was fulfilled. The system consists of a semi-infinite Si-substrate, a 6 nm thick layer of n-type doped $Bi_2Te_3$ and a layer of varying thickness of p-type doped $Sb_2Te_3$ (Figure 5, inset). The electron and hole effective masses as well as the band gaps were taken from the results of ab-initio calculations [3, 4]. To our knowledge, the band offset between Si and $Bi_2Te_3$ has so far not been determined and was assumed to be equal to the difference between the electron affinities of Si (4.05 eV from ref. [5]) and $Bi_2Te_3$. The $Bi_2Te_3$ electron affinity ranges from 4.125 to 4.525 eV (ref. [6]), thus the band offset value of 0.3 eV between Si and $Bi_2Te_3$ seems acceptable. Values for the dielectric constant for both materials were found in literature to be $\varepsilon_{Bi_2Te_3} = 75$ [7] and $\varepsilon_{Sb_2Te_3} = 36.5$ [8]. Different numbers for the bulk native defects and surface states density were computed and the resulting band diagrams compared to experimental spectra. The values with best agreement were found to be for donor-type native defects in $Bi_2Te_3$ - $N_d = 2 \cdot 10^{19}$ cm$^{-3}$ and acceptor-type native defects in $Sb_2Te_3$ - $N_a = 2 \cdot 10^{18}$ cm$^{-3}$. Additionally, two layers of negative surface charge with a density of $1 \cdot 10^{12}$ cm$^{-2}$ were assumed at the interfaces between topologically trivial to non-trivial materials, i.e. interface of $Bi_2Te_3$ to the Si substrate and at the $Sb_2Te_3$ surface to vacuum. All these charge densities are within experimentally confirmed uncertainty limits. All calculations were performed assuming $T$ = 20 K. Instead of an abrupt defect distribution step, an intermixed interface region of 5 nm width derived from Supplementary Fig. 1b was included in the model. The band gap within the two materials has been linearly changed from 0.12 eV for $Bi_2Te_3$ to 0.16 eV for $Sb_2Te_3$ within this region. The model calculates the band diagram through the entire sample but only the resulting valence band position at the surface to vacuum was plotted against top layer $Sb_2Te_3$ thickness and compared to experimental ARPES results in Figure 5.

**Estimations for extreme limits of the model.**

In order to evaluate the effect of intermixing on the position of the energy bands, we modeled also for two extreme cases of no and very strong intermixing. Supplementary Fig. 2 again shows the position of the valence band maximum vs. $Sb_2Te_3$ top layer thickness. The red dots and the black curve are the same as in Figure 5 and show experimental ARPES data and the results of the model for an intermixing which is based on the measured profile in Supplementary Fig. 1b. Additionally, the red curve depicts the results of the model for a perfectly sharp interface, i.e. no intermixing at all, and the green curve for strong gradual intermixing of the charge carrier density through the entire structure (the carrier density distribution is depicted on the right of Supplementary Fig. 2 for each curve). As one can see, the green curve is quite different from our experimental results, whereas the red and black curve are similar, showing that strong intermixing would not match the ARPES and transport results.

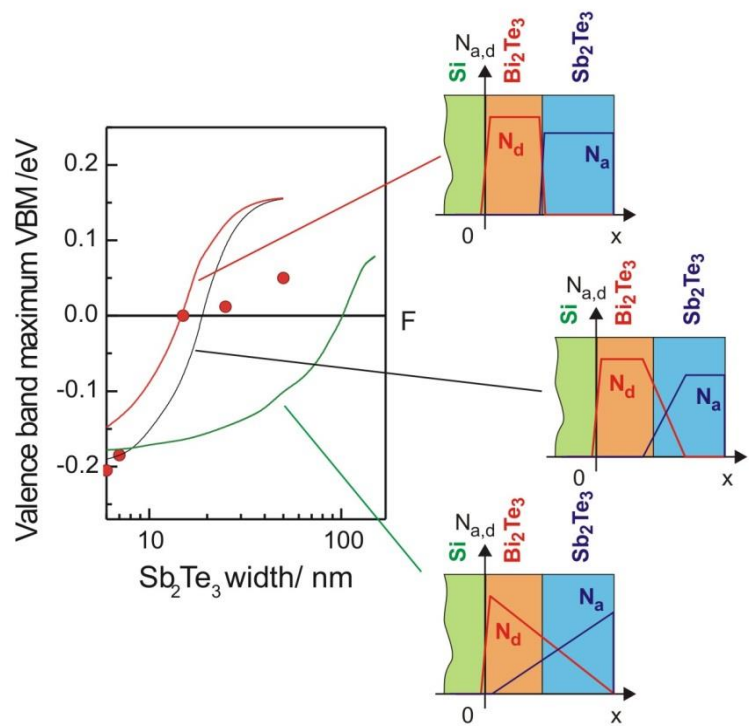

**Supplementary Figure 2 | Estimation of the limits of our 1D model.** The calculated energetic position of the valence band maximum is plotted against $Sb_2Te_3$ top layer thickness. Experimental results (red dots) are compared to the results from our 1D Schrödinger Poisson model calculations. Red curve shows result for a perfectly sharp interface between the two $Bi_2Te_3$ and $Sb_2Te_3$ layers and the green curve, respectively, for a very strong gradual intermixing of charge carriers. The black

curve assumes a light composition profile on the basis of the experimentally deduced profile (from Supplementary Fig. 1b). Each situation is schematically illustrated. Green is the Si substrate, orange the $\text{Bi}_2\text{Te}_3$ layer whose thickness is fixed and blue the varying $\text{Sb}_2\text{Te}_3$ layer. Red and blue lines here represent donor- and acceptor-type carrier distribution, respectively.